\documentclass[%
 reprint,
twocolumn,
superscriptaddress,
 amsmath,amssymb,
 aps,
prd,
preprintnumbers,
]{revtex4-1}

\usepackage{graphicx}
\usepackage{dcolumn}
\usepackage{bm}
\usepackage{aas_macros}
\usepackage{hyperref}
\usepackage{xcolor}
\usepackage[mathlines]{lineno}

\usepackage{amsmath}
\usepackage{comment}
\usepackage{xspace}

\newcommand{\mwdm}{m_{\rm WDM}\xspace}
\newcommand{\fwdm}{f_{\rm WDM}\xspace}
\newcommand{\fnu}{f_{\nu_s}\xspace}

\newcommand{\CYone}[1]{{#1}}
\newcommand{\CYtwo}[1]{{#1}}

\begin{document}

\preprint{FERMILAB-PUB-24-0586-PPD}

\title{Mixed Warm Dark Matter Constraints using Milky Way Satellite Galaxy Counts}

\author{Chin Yi Tan}
\email{chinyi@uchicago.edu}
\affiliation{Kavli Institute for Cosmological Physics, University of Chicago, Chicago, IL 60637, USA}%
\affiliation{Department of Physics, University of Chicago, Chicago IL 60637, USA}
 
\author{Ariane Dekker}%
 
\affiliation{Kavli Institute for Cosmological Physics, University of Chicago, Chicago, IL 60637, USA}%

\author{Alex Drlica-Wagner}
\affiliation{Fermi National Accelerator Laboratory, P. O. Box 500, Batavia, IL 60510, USA}
\affiliation{Kavli Institute for Cosmological Physics, University of Chicago, Chicago, IL 60637, USA}
\affiliation{Astronomy \& Astrophysics, University of Chicago, Chicago IL 60637, USA}

\date{\today}

\begin{abstract}
Warm dark matter has been strongly constrained in recent years as the sole component of dark matter. However, a less-explored alternative is that dark matter consists of a mixture of warm and cold dark matter (MWDM). In this work, we use observations of Milky Way satellite galaxies to constrain MWDM scenarios where the formation of small-scale structure is suppressed either by generic thermal relic warm dark matter or a sterile neutrino produced through the Shi-Fuller mechanism. To achieve this, we model satellite galaxies by combining numerical simulations with semi-analytical models for the subhalo population, and use a galaxy--halo connection model to match galaxies onto dark matter subhalos. By comparing the number of satellites predicted by  MWDM models to the observed satellite population from the Dark Energy Survey and Pan-STARRS1, we constrain the fraction of warm dark matter, $\fwdm$, as a function of its mass, $\mwdm$. We exclude dark matter being composed entirely of thermal relic warm dark matter with  $\mwdm \leq 6.6 $ keV at a posterior ratio of 10:1, consistent with previous works. However, we find that warm dark matter with smaller mass is allowed when mixed with cold dark matter, and that the $\fwdm$ constraints strengthen with decreasing $\mwdm$ until they plateau at $\fwdm \lesssim 0.45 $ for $\mwdm \lesssim 1.5$ keV. Likewise, in the case of a sterile neutrino with mass of 7~keV produced through the Shi-Fuller mechanism, we exclude a fraction of $f_{\nu_s} \lesssim 0.45$, independent of mixing angle. Our results extend constraints on MWDM to a region of parameter space that has been relatively unconstrained with previous analysis. 
\end{abstract}

\maketitle


\section{Introduction} \label{sec:intro}
The existence of non-baryonic dark matter can be inferred from a variety of astronomical and cosmological observations accumulated over the last century (see reviews by \citet{Einasto:2009} and \citet{Bertone:2018}). However, its particle nature still remains unknown and is considered to be one of the most important open problems in modern cosmology and particle physics, as shown by the enormous investment in experimental and theoretical investigations of dark matter (e.g., see reviews by \citet{Bertone:2005} and \citet{Freese:2017}).

The current cosmological paradigm consists of a cosmological constant, $\Lambda$, and collisionless cold dark matter (CDM). In this model, dark matter is composed of a heavy, stable and non-interacting  particle responsible for structure formation~\citep{Peebles:1982, Blumenthal:1984}. 
While  observations of the cosmic microwave background (CMB), big bang nucleosynthesis (BBN), galaxy clustering, weak lensing, and the Lyman-$\alpha$ forest are in good agreement with the CDM model at length scales of $\gtrsim 0.1$ Mpc~\citep{Chabanier:2019, Planck:2020, Ferreira:2021}, smaller scales are less observationally constrained, and dark matter might have properties that deviate from the standard CDM model at these smaller scales \citep[e.g.,][]{Bullock:2017}.

Due to their high dark matter content, small size and close proximity, the ultra-faint dwarf galaxy (UFD) satellites of the Milky Way (MW) are excellent systems to probe the properties of dark matter at the smallest scales \citep{Simon:2019}. For example, the properties of dark matter such as its mass, interaction cross section, and thermal history can have a large impact on the number counts, luminosity function, kinematics and density profiles of UFDs  \citep[e.g.,][]{Lovell:2014, Rocha:2013, Kaplinghat:2015, Penarrubia:2016, Bullock:2017, Nadler:2019b, Dalal:2022, Esteban:2023}. UFDs are also among the best targets for indirect searches for dark matter annihilation or decay products (see \citet{Strigari:2018} for a review).

One alternative to the CDM model is warm dark matter (WDM), which has free-streaming effects that can suppress structure formation and thus the formation of dwarf galaxies \citep{Kolb:1990}. One possible WDM candidate is the sterile neutrino \citep{Dodelson:1994, Shi:1999} with right-handed chirality and no electric charge, which interacts with Standard Model particles through mixing with either neutrinos or through some beyond the Standard Model interaction \citep{Abazajian:2017, Adhikari:2017, Boyarsky:2019}.
The claimed detection of a 3.57 keV X-ray line in the halos of galaxies and galaxy clusters \citep{Boyarsky:2014, Bulbul:2014} drew particular attention to a sterile neutrino with mass of $\sim$7 keV. However, follow-up observations and reanalysis have not reproduced the detection of this X-ray line~\citep{Slatyer:2017, Dessert:2020, Dessert:2024}. 


Numerous studies place strict constraints on WDM as the sole component of dark matter for both the general thermal relic case and specific sterile neutrino models \citep{Zelko:2022, Lovell:2023}. 
Using MW satellite counts, various works have placed stringent constraints on the mass of thermal relic WDM~\citep{Kennedy:2013uta,Lovell:2015psz,Enzi:2020ieg,Newton:2020cog,Dekker:2022,Newton:2024}, the strongest of which yields a 95\% Bayesian lower limit of $\mwdm \geq$ 6.5~keV \citep{Nadler:MW3}. 
Other measurements of the Lyman--$\alpha$ forest~\citep[e.g.,][]{Irvsivc:2024}, UV luminosity function~\citep[e.g.,][]{Liu:2024}, stellar streams~\citep[e.g.,][]{Banik:2021} and gravitationally lensed quasars \citep[e.g.,][]{Keeley:2024} have also been used to set limits on the thermal relic WDM mass at $\mwdm \geq$ 5.7, 3.2, 3.6, and 6.1 keV, respectively (using a variety of significance criteria). Moreover, through a combined analysis of MW satellites counts and \CYtwo{flux ratios of  quadruply-lensed quasars}, \citet{Nadler:2021} produced a stringent joint constraint of $\mwdm \geq$ 9.7~keV at a  95\% Bayesian lower limit and 7.4~keV at a 20:1
marginal likelihood ratio. While \citet{Enzi2021} obtained limits of 
$\mwdm \geq$ 6.0~keV at a  95\% Bayesian lower limit and 2.6~keV at a 20:1 likelihood ratio by combining the analysis of \CYtwo{resolved imaging of galaxy--galaxy lenses},  Lyman-$\alpha$ forest, and MW satellite counts.

Such constraints are alleviated if dark matter is composed partly of WDM and partly of CDM
~\citep{Anderhalden:2012, Parimbelli:2021}. Compared to the pure WDM scenario, there has been relatively little work constraining mixed warm dark matter (MWDM) scenarios. Using 21-cm absorption lines, \citet{Schneider:2018} found that the WDM fraction cannot be larger than 17 percent of the total dark matter abundance for thermal-relic WDM  with mass of $\mwdm \leq 1$ keV. Furthermore, \citet{Boyarsky:2009} used the Lyman--$\alpha$ forest to investigate the mixed CDM and sterile neutrino scenario, allowing sterile neutrino fraction below 60 percent of the total dark matter abundance for non-resonantly produced sterile neutrino with mass of $m_{\nu_s} \geq 5$ keV. Additionally, \citet{Inoue23} and \citet{Keeley:2023} forecast future prospects for constraining MWDM models using JWST observations of lensed quasars. Many of these constraints are made with different model assumptions, therefore it is important to have complementary searches.

In this analysis, we \CYtwo{compare predictions for the luminosity functions of MW satellite galaxies from various MWDM models with the MW satellite galaxies observed in  the Dark Energy Survey \citep[DES:][]{DES:2016} and the Pan-STARRS1 \citep[PS1:][]{Chambers:2016}  survey to set constraints on MWDM properties. } In Section \ref{sec:sashimi}, we first discuss the Semi-Analytical Sub-Halo Inference ModelIng code \citep[\texttt{sashimi};][]{Hiroshima_2018, Ando_2019, Dekker:2022} used to obtain the mass distribution of subhalos for different MWDM scenarios and present an analytical approximation for the suppression of the subhalo mass function relative to CDM. 
Section \ref{sec:constraints} describes the galaxy--halo connection model \citep{Nadler:MW3} used to predict the number of MW satellites that would be observable in different MWDM scenarios. We  compare these predictions to the observed MW satellite counts to obtain constraints on mixed thermal-relic scenarios (Section \ref{sec:limits_thermalWDM}) and mixed sterile neutrinos scenarios (Section \ref{sec:limits_sterilenu}). We discuss our analysis and conclude in Section \ref{sec:summary}. In our analysis, we adopt the flat $\Lambda$MWDM model as our fiducial cosmology with $H_0 = 70$ km s$^{-1}$ Mpc$^{-1}$, $h = 0.7$, $\Omega_{\rm m} = 0.286$, $\Omega_{\Lambda} = 0.714$, $\Omega_{\rm b} = 0.047$ , $\sigma_8 = 0.82 $, $n_s = 0.96 $.

\section{Subhalo population of Mixed Thermal-relic WDM} \label{sec:sashimi}
The free-streaming length of WDM suppresses the formation of low-mass subhalos that could host the faintest galaxies~\citep{Lovell:2014}. However, this suppression is reduced in the presence of CDM. In this work, we consider mixed dark matter with total dark matter abundance given by $ \Omega_{\rm{DM}} = \Omega_{\rm{WDM}} + \Omega_{\rm{CDM}}$, and a WDM fraction given by  $\fwdm = \Omega_{\rm{WDM}} / \Omega_{\rm{DM}}$. 

In our analysis, we use the subhalo populations obtained from two \texttt{GADGET-2}-based \citep{Springel:2005} cosmological zoom-in simulations from \citet{Mao15} as a basis for the CDM subhalo mass function. We have selected two simulated host halos that have similar masses ($M = 1.57$ and $1.26 \times 10^{12}M_\odot$), concentrations ($c = 11.8$ and 10.5.), and assembly histories as the MW halo, as well as an analog of the  Large Magellanic Cloud halo. The simulations  have a minimum particle mass of $3\times10^5$ $M_\odot h^{-1}$.

We account for the MWDM scenario by multiplying the distribution of subhalo masses in CDM, known as the subhalo mass function,
with a subhalo suppression function, \CYtwo{$\beta(M,\theta_{\rm WDM},\fwdm)$, such that}  
\begin{equation}\label{eq:beta}
    \left(\frac{dN_ {\rm sub}}{dM} \right)_{\rm MWDM} =  \beta(M,\theta_{\rm WDM},\fwdm) \left(\frac{dN_ {\rm sub}}{dM} \right)_{\rm CDM},
\end{equation}
where $dN_{\rm sub}/{dM}$ is the subhalo mass function, \CYone{ $M$ is the subhalo mass,  \CYtwo{$\theta_{\rm WDM}$} is the model-specific \CYtwo{parameters} of the WDM scenario such as its particle mass, and $\fwdm$ is the WDM fraction. Throughout this work, we calculate the subhalo suppression function based on the subhalo mass at accretion within a radius where its mean density is 200 times the critical density of the Universe, which we assume is equal to its peak mass.}

We further assume that the only difference between the MWDM and CDM scenario is the number of subhalos, and all other properties remain identical. 
Indeed, it has been demonstrated that the radial distribution of the subhalos large enough to consistently form galaxies does not change between CDM and WDM cosmologies \citep{Lovell:2014, Bose:2017, Lovell:2021}.
However, other differences between the CDM and MWDM models may affect the abundance of MW satellite galaxies. For instance the concentration-mass-redshift relation is different for CDM and WDM~\citep{Ludlow:2016, Gilman:2020}, which might further reduce the predicted number of MW satellites. Our results are conservative in the sense that we do not account for this reduction.

In order to obtain the suppression function for a large range of \CYone{WDM scenarios}, we use the semi-analytical \texttt{sashimi} code \citep{Hiroshima_2018, Ando_2019, Dekker:2022}. 
\texttt{sashimi} predicts subhalo properties 
for a given matter power spectrum and host halo mass, 
and has been tested against numerical simulations in the case of pure CDM and WDM. \CYtwo{Specifically for this work, we have verified that the CDM subhalo mass function generated by \texttt{sashimi} is consistent with the
subhalo populations obtained from the two simulations by \citet{Mao15}.} Semi-analytical models provide a fast and flexible platform for computing subhalo properties, allowing us to make quick estimates of the predictions from different dark matter scenarios. Alternative approaches exploring MWDM scenarios through cosmological N-body simulations are forthcoming and will provide a valuable cross check \citep{An:2024}.

Here, we describe the case of MWDM comprised of a thermal relic WDM candidate \CYone{ with a particle mass of $\mwdm$}, and we discuss applications to a sterile neutrino WDM candidate in Section~\ref{sec:limits_sterilenu}. 
We compute the matter power spectrum with the Boltzmann code \texttt{CLASS}~\citep{lesgourgues2011cosmic}, where we specify $m_{\text{WDM}}$, $f_{\text{WDM}}$, and the relative temperature of WDM with respect to the photon temperature, $T_X/T_{\gamma}$. 
We obtain the relative temperature through
\begin{equation}
    \frac{T_{\text{X}}}{T_{\gamma}} = \left( \frac{4}{11}\right)^{1/3} \left( \frac{43/4}{g_{\star}^{dec}}
    \right)^{1/3},
\end{equation}
where $g_{\star}^{dec}$ is the spin degrees of freedom at decoupling, which can be found through the WDM abundance as
\begin{equation}
    \Omega_{\rm WDM} h^2 \approx \frac{115}{g_{\star}^{dec}} \frac{g_X}{1.5} \frac{m_{\text{WDM}}}{\text{keV}},
\end{equation}
with $g_{X}=1.75$ the degrees of freedom for a fermionic dirac dark matter ~\citep{Bode:2000,Vogel:2022}.
Moreover, when computing the matter power spectrum in \texttt{CLASS}, we turn off the fluid approximation through \textsc{ncdm\_fluid\_approximation~$=3$} to increase precision.  We observe that the transfer functions reach a plateau at large $k$ due to the CDM component at a height that depends on $\fwdm$.

We obtain the subhalo suppression functions (Eq.~\ref{eq:beta}) with \texttt{sashimi} for host halo mass of $10^{12}M_\odot$ \footnote{\CYone{The subhalo suppression function remains unchanged when using the simulated host halo mass value of $1.57 \times 10^{12}M_\odot$.}} and consider $m_{\rm{WDM}}=(1.5-8.0)$~keV with interval of 0.5~keV, and $\fwdm=(0.1-1.0)$ with interval of 0.1. In order to obtain an analytical expression for the suppression function, we fit the \CYone{mixed thermal-relic WDM} subhalo suppression functions to the following analytical function
\begin{equation}
      \beta(M,\mwdm,\fwdm) \approx \Delta \left [ 1+ \alpha \left(  \frac{10^9M_\odot}{M}\right) \right]^{\gamma} + (1- \Delta),
\label{eqn:beta_fit}
\end{equation}
where $\alpha$, $\Delta$, $\gamma$ are  fitting parameters. We find a good fit with the \texttt{sashimi} suppression functions when we parameterize the fitting parameters as:
\begin{equation}
\label{eqn:alphagammadelta}
\begin{aligned}
 \alpha &= 29.5 ({\mwdm}/ 1 \rm{keV}) ^{-3.1} \\
 \gamma &=  -0.980 \fwdm^2 -0.096 \fwdm -0.203  , \\
 \Delta &= \Delta_0 \fwdm^2 + (1-\Delta_0) \fwdm . 
\end{aligned}
\end{equation}
where $\Delta_0$ is given by 
\begin{equation}
\Delta_0 = \frac{-1}{1+\exp(-1.358((\mwdm/\rm{keV})  - 0.80 ))}
\end{equation}

\noindent For comparison, \citet{Lovell:2014} found the following fitting parameters in the case of pure WDM: \CYtwo{$\alpha \approx 52.6 (\mwdm/1 \rm {keV})^{-3.33}$, $\gamma = -0.99$, and $\Delta = 1$, assuming that the half-mode mass is given by $M_{\rm hm} = 5 \times 10^8  (\mwdm/3 \rm {keV})^{-10/3} M_\odot $\citep{Nadler:MW3}.} 

\begin{figure*}[t!]
\centering
\includegraphics[width=0.99\linewidth]{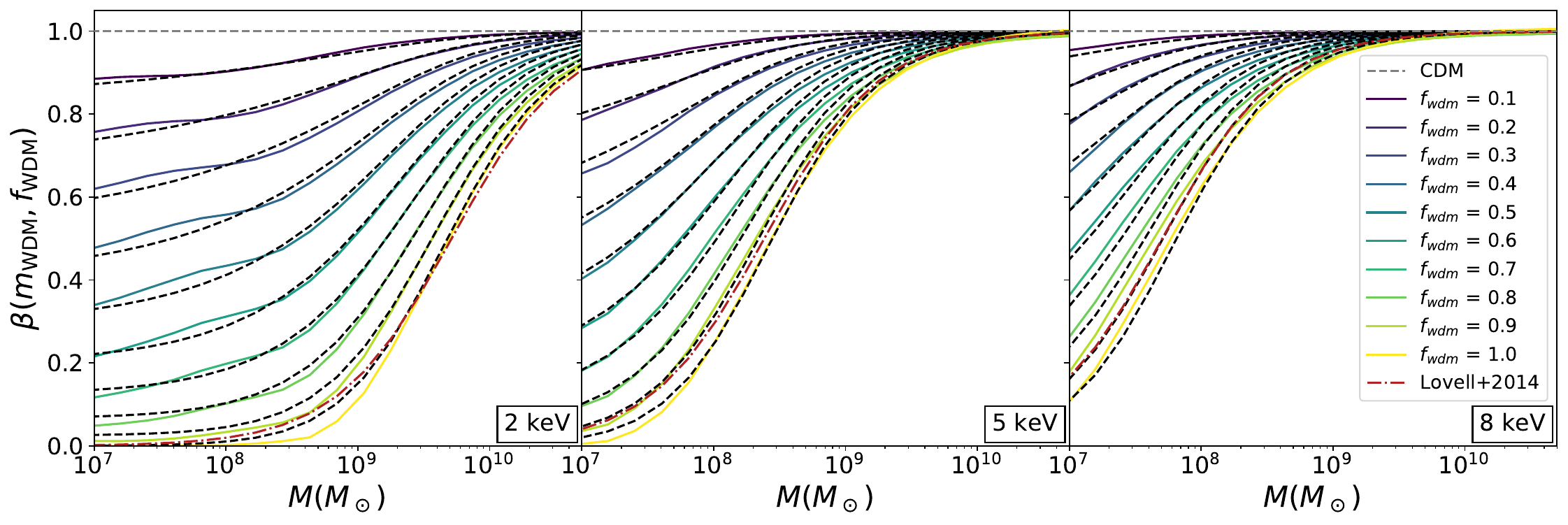}
\caption{The subhalo suppression function, $\beta(M,\mwdm,\fwdm)$, for thermal-relic WDM with mass of 2 keV (left), 5 keV (middle), 8 keV (right) as a function of subhalo mass $M$. The colored solid lines show the  suppression functions for different fractions, $\fwdm$, from \texttt{sashimi}, while the black dashed lines show the best-fit analytical suppression functions (see Eq. \ref{eqn:beta_fit}). \CYtwo{Also shown is the pure thermal-relic WDM ($\fwdm$=1) subhalo suppression function from \citet{Lovell:2014} }
\label{Figure:f_wdm} }
\end{figure*}

Fig.~\ref{Figure:f_wdm} illustrates the subhalo suppression functions obtained from the analytical fits (dashed) compared to the subhalo suppression functions obtained with \texttt{sashimi} (solid) for WDM mass of 2~keV (left), 5~keV (middle) and 8~keV (right). The difference between the two functions is less than 5\% for subhalo masses in the range of $10^{7} < M/M_\odot < 10^{10.5}$ and thermal-relic WDM mass of 1.5 keV $\leq \mwdm \leq$ 8keV. We find that our analytical fitting functions break down for $\mwdm < 1.5$ keV, and thus we do not include those low masses in our analysis.

\section{galaxy--halo connection model} \label{sec:constraints}

We adopt the galaxy--halo connection model described in \citet{Nadler:2019, Nadler:MW2} to predict the number of MW satellite galaxies that are expected to be observed in DES and PS1 for different MWDM scenarios. The model  ``paints'' satellite galaxies onto the well-resolved simulated subhalos with a peak mass of  $M>10^7 M_\odot$, present-day maximum circular velocity of $V_{\rm max}>$ 9 km s$^{-1}$ and peak circular velocity of $V_{\rm peak} >$~10 km s$^{-1}$, as discussed in the following.


In this galaxy--halo connection model, the total number of predicted MW satellites, $n_{\rm sat}$, is given by 
\begin{equation}
\label{eqn:subhalo_number}
    n_{\rm sat} = \sum_{i}   f_{{\rm gal}, i}\times (1 - p_{{\rm disrupt}, i}) \times p_{{\rm detect}, i} \times  \beta_{i}
\end{equation}
where $i$ indexes the simulated subhalos,  $f_{\rm gal}$ is the galaxy occupation fraction (probability that the subhalo will host a satellite galaxy), $ p_{\rm disrupt}$ is the probability that a subhalo would be disrupted due to baryonic effects \citep{Nadler:2018}, $p_{\rm detect}$ is probability of the  galaxy being detected in DES and PS1 calculated using  survey selection functions from \citet{Drlica-Wagner:MW1}, and $\beta$ (Eq.~\ref{eqn:beta_fit}) is the subhalo suppression function relative to the  simulated subhalo population in CDM. For a given set of galaxy--halo connection model parameters ($\theta_{\rm model} = \{ \alpha,\sigma_M, \mathcal{A}, n, \sigma_{\log R}, \mathcal{M}_{50}, \sigma_{\rm gal}, \mathcal{B}\}$, see  \citet{Nadler:MW2} for more details), and a specific subhalo suppression function, $\beta$, we can generate a predicted population of satellite galaxies.


\section{Limits on Mixed Thermal Relic WDM} \label{sec:limits_thermalWDM}
To constrain MWDM scenarios, we compare the predicted  galaxy counts from individual MWDM models with the observed counts of 17 and 19 known satellites from DES and PS1, respectively (see \citet{Nadler:MW2} for details on how the satellites are selected). 

To incorporate the luminosity and size information of the satellites, we bin the predicted and observed satellites based on their absolute magnitude and surface brightness \footnote{We first bin the galaxies into \CYtwo{6} equally-spaced  bins based on their absolute magnitude from $0 \geq M_V \geq -20$. We further split the galaxies into 4 bins, based on a surface brightness threshold at 28 mag arcsec$^{-2}$ and whether the galaxies would be found in the DES and PS1, resulting in a total of \CYtwo{24} galaxy bins.}. We obtain the likelihood that the predicted number of galaxies for each MWDM scenario represent the observed MW satellite population in DES and PS1 by assuming that the predicted number of satellite galaxies follows a Poisson distribution in each bin \citep{Nadler:MW2}. 

\begin{figure}[h]
\centering
\includegraphics[width=\linewidth]{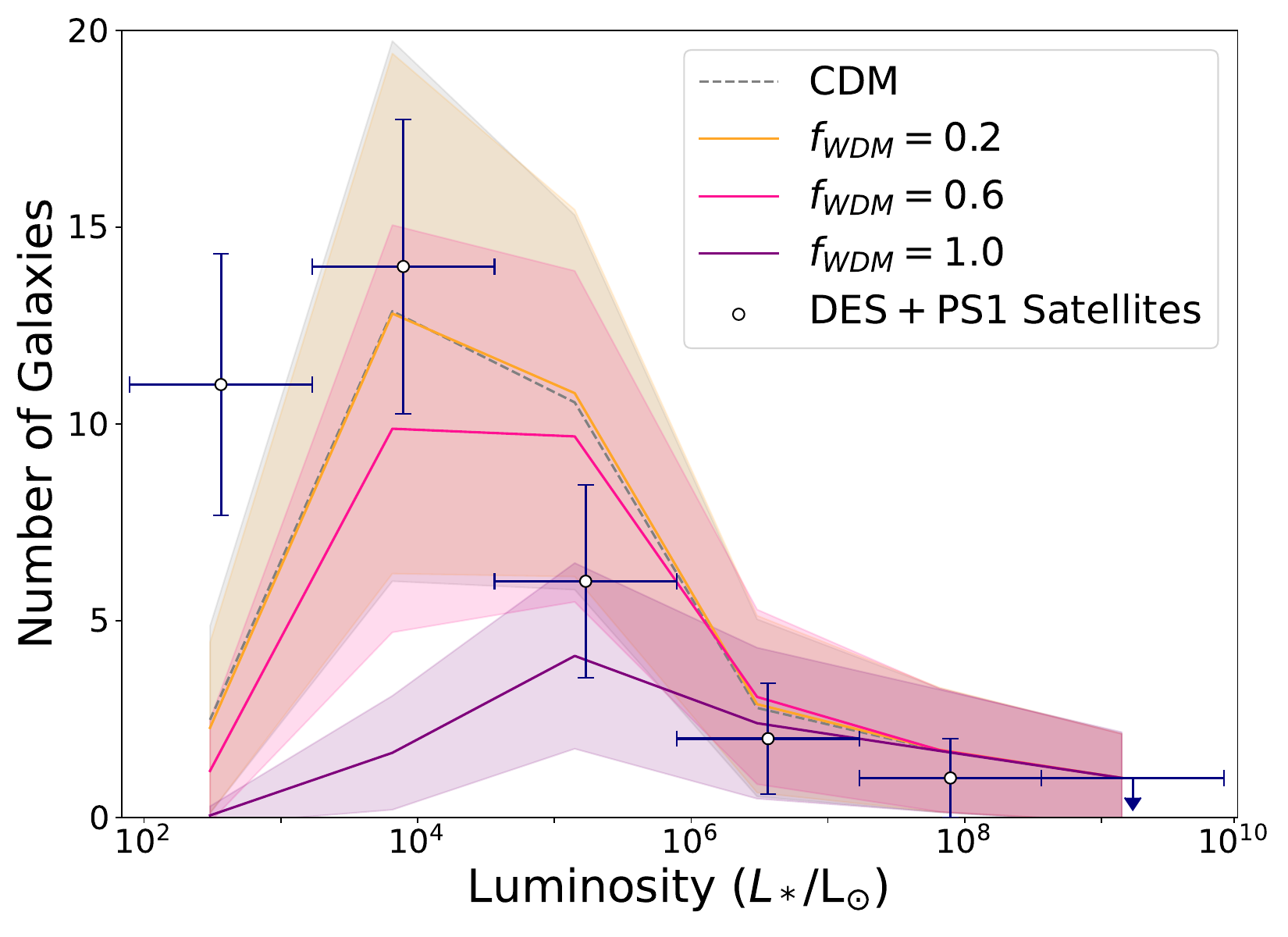}
\caption{Predicted luminosity functions of observable MW satellite galaxies in DES and PS1 for CDM (grey dashed line) and MWDM scenarios with $\mwdm = 2$\,keV and various $\fwdm$ (colored lines), compared to the observed number of galaxies (error bars). The shaded bands represent the uncertainty due to different  galaxy--halo connection model values and the stochasticity of the model. 
\label{Figure:numgal}  }
\end{figure}

Fig. \ref{Figure:numgal} shows the binned predicted luminosity functions of galaxies observed in both the DES and PS1 regions for CDM and MWDM \CYone{comprising of thermal-relic WDM} with $\mwdm = 2$\,keV.  For each scenario, we use the Markov Chain Monte Carlo sampler \texttt{emcee} \citep{Foreman_Mackey:2013} to sample the posterior distribution of the galaxy--halo connection model parameters. The shaded band represents the uncertainty due to different  galaxy--halo connection model values and the stochasticity of the model. The uncertainties of the luminosities and number counts of the observed satellites is associated with the size of the luminosity bins used in the analysis and Poisson statistics, respectively. We find that due to the uncertainty in the  galaxy--halo connection model parameters, the $\fwdm=0.2$, $\mwdm = 2$\,keV \CYone{mixed thermal-relic WDM scenario} predicts roughly the same number of MW satellite galaxies as CDM, while the $\fwdm=0.6$ scenario starts to show a deviation from CDM. Note that for illustrative purposes,  we only bin the galaxies based on their luminosities in Fig. \ref{Figure:numgal}.

We use \texttt{emcee} \citep{Foreman_Mackey:2013} to sample the marginalized posterior distribution of the galaxy--halo connection model for different \CYone{mixed thermal-relic WDM } models. In our analysis, we scan over fixed WDM mass and leave the WDM fraction as a free parameter between $0 \leq \fwdm \leq 1$. For each WDM mass, we run \texttt{emcee} for 20,000 steps with 8 walkers to sample the posterior of nine free parameters in the model, $ \theta_{\rm mcmc} =  \{ \alpha,\sigma_M, \mathcal{A}, n, \sigma_{\log R}, \mathcal{M}_{50}, \sigma_{\rm gal}, \mathcal{B}, \fwdm  \}$. The first eight parameters are the galaxy--halo connection model parameters, and we marginalize over them in this analysis. 
We run the MCMC chains for WDM masses in the range of $1.5 \leq \mwdm \leq 8.0$\,keV at mass intervals of 0.5 keV. 

\begin{figure}[t]
\centering
\includegraphics[width=\linewidth]{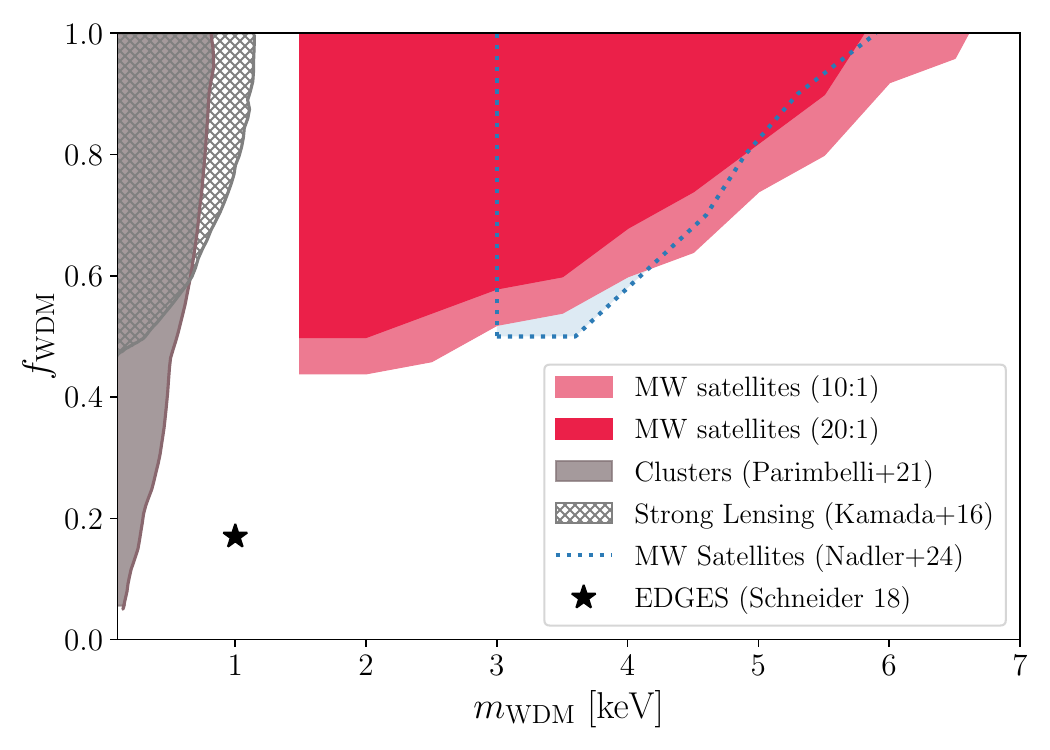}
\caption{\label{Figure:constrained}
The exclusion region for 
WDM mass, $\mwdm$, as a function of WDM fraction, $\fwdm$,  
with posterior ratio threshold of 10:1 (light pink) and 20:1 (dark pink).  The grey regions represent the parameter space that have been excluded by previous analyses using strong gravitational
lenses \citep{Kamada:2016} and cluster number counts \citep{Parimbelli:2021}. \CYone{While the blue region represents parameter space excluded by MW satellite observations with WDM subhalos obtained from  cosmological N-body zoom-in simulations \citep{An:2024}.} The star represent constraints from  21-cm absorption feature detected by EDGES~\citep{Schneider:2018}. 
}
\end{figure}

To obtain constraints, we find the value of $\fwdm$ that yields a posterior odds ratio compared to CDM  of  $\mathcal{P}_{\rm CDM}$\,:\,$ \mathcal{P}_{\rm MWDM} = 10$:$1$ (and  20:1) following the approach of \citet{Nadler:2021} and \citet{Keeley:2024}. We  obtain the marginalized posterior odds ratio, $\mathcal{P}_{\rm MWDM}$\,:\,$ \mathcal{P}_{\rm CDM}$, by comparing the number of walkers for particular WDM fraction, with the walkers at the CDM limit $\fwdm=0$ using a histogram with bin size of $\fwdm=0.02$. 

An alternative method commonly quoted in the WDM literature is the 95\% credible interval \citep{Viel:2013, Nadler:MW3},  which is defined as the range of parameter values which enclosed 95\% of the integrated marginal posterior. While the credible interval methods includes more information of the shape of the posterior, our posteriors are unbounded in the limit of large $\mwdm$, and thus the limits obtained would be sensitive to the boundary of the  priors adopted for the WDM parameters which can be arbitrary \footnote{The full posterior distribution can be found  \url{https://github.com/chinyitan/mwdm}.}.

Fig.~\ref{Figure:constrained} shows the parameter space that is disfavored by the MW satellite galaxy counts. We exclude \CYone{thermal-relic} WDM with mass $\mwdm \lesssim 6.6$ keV (5.8 keV) with a posterior ratio of 10:1 (or 20:1)  from being all of dark matter. This is  similar to the pure WDM constraints found in \citet{Nadler:2021}, which excluded $\mwdm \leq  5.2$ keV  at marginalized posterior ratio of 20:1 using a slightly different form for the suppression of the subhalo mass function. As we move towards lower $\mwdm$, our allowed  WDM fraction, $\fwdm$, decreases until our constraints starts to converge at $\fwdm \sim 0.45$ (0.50).  

In Fig.~\ref{Figure:constrained}, we also compare our constraints with previous constraints, as shown as grey shaded regions from gravitational lenses~\citep{Kamada:2016}, and cluster number counts \citep{Parimbelli:2021}. \CYone{Additionally, a similar analysis constraining MWDM models by comparing MW satellite observations to cosmological N-body zoom-in simulations yields  qualitatively similar results \citep{An:2024}, lending strength to both approaches.} Moreover, the star represent constraints from the  21-cm absorption feature claimed by EDGES \citep{Bowman:2018}, which excludes $\fwdm \geq 0.17$ for $\mwdm \sim$ 1 keV \citep{Schneider:2018}. 

\CYtwo{In this analysis, we adopted the same wide priors on the galaxy--halo connection model parameters as \citet{Nadler:MW2}. As a result, in some MWDM scenarios, the suppression in the subhalo mass function can be compensated by adjusting the values of the galaxy--halo connection model parameters. This, in turn, weakens our MWDM constraints.  For example, we find that the posterior distribution of the faint-end slope of the subhalo-satellite luminosity function, $\alpha$, in our MWDM scenarios is lower compared to CDM. The faint-end slope determines the luminosity of a satellite galaxy as a function of subhalo peak circular velocity. A lower value of $\alpha$ allows less massive subhaloes to produce brighter satellites, making these galaxies more easily detectable and consequently increasing the predicted number of detected galaxies. We also observe similar effects in the other galaxy--halo connection model parameters but to a much smaller extent. A more extensive discussion of the galaxy--halo connection model can be found in \citet{Nadler:2019} and \citet{Nadler:MW2}}


\begin{figure}[t!]
\centering
\includegraphics[width=\linewidth]{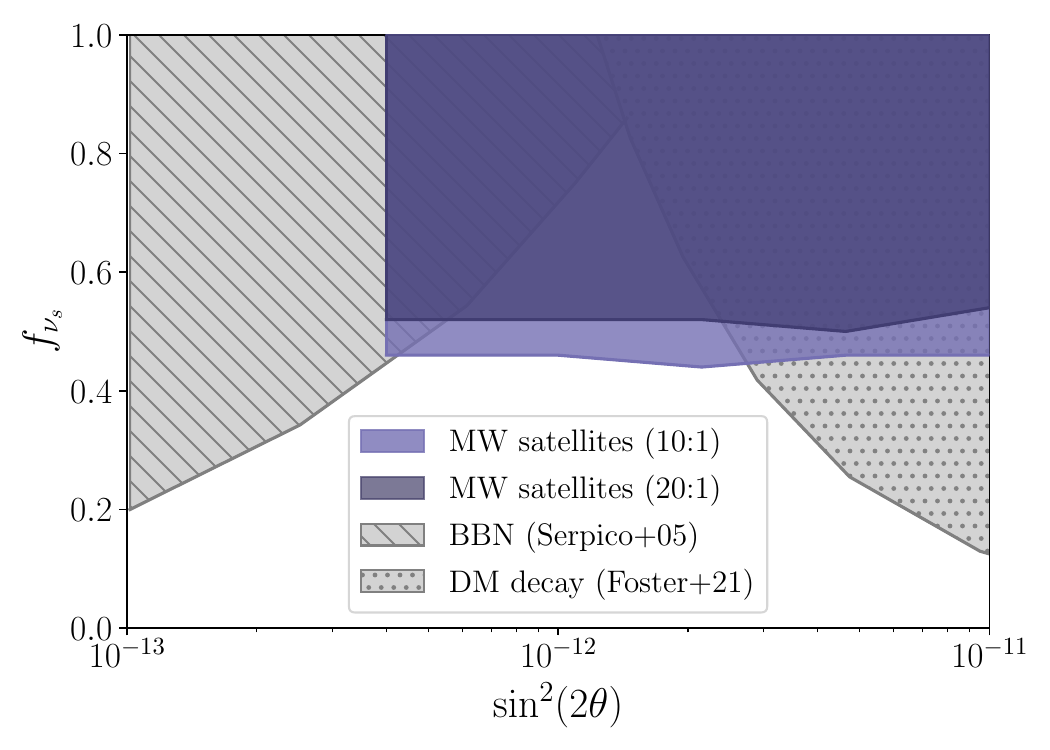}
\caption{
The exclusion region for 7~keV sterile neutrino mixing angle, $\sin^2(2\theta)$, as a function of its fraction, $f_{\nu_s}$ with posterior ratio threshold of 10:1 (light purple) and 20:1 (dark purple). The grey dotted region comes from X-ray constraints with XMM-Newton \citep{Foster:2021ngm}  and the grey hatched region comes from limits from BBN \citep{Serpico:2005bc}. 
\label{Figure:sterile_nu_constrained}
}
\end{figure}

\section{Limits on Mixed Sterile Neutrino }  \label{sec:limits_sterilenu}
We apply our analysis to a MWDM model consisting of a $m_{\nu_s}=7$\,keV sterile neutrino produced through the Shi-Fuller mechanism~\citep{Shi:1999}, in which its production is resonantly enhanced due to the presence of a net lepton asymmetry in the primordial plasma. 
Sterile neutrino's are not produced thermally, and their spectral energy distributions are typically warmer than thermal WDM at comparable mass.

We use the public \texttt{sterile-dm} code \citep{Venumadhav:2016} to generate the phase-space distributions of $\nu_s$ and $\bar{\nu}_s$ for a range of lepton asymmetry values that correspond to sterile neutrino abundances, $\Omega_{\nu_s}$, and mixing angles, $\sin^2(2\theta)$.
We obtain the shape of the matter power spectrum by providing the phase-space distributions as an input for the Boltzmann code \texttt{CLASS}~\citep{lesgourgues2011cosmic}. 
We obtain the subhalo suppression function, \CYtwo{$\beta(M,\sin^2(2\theta),f_{\nu_s})$,} using \texttt{sashimi} for mixing angles and fractions in the range of  $  \CYone{4\times10^{-13}} \leq \sin^2(2\theta) \leq 10^{-11}$ and $0.1 \leq f_{\nu_s} \leq 0.6$ with fraction steps of 0.1. The subhalo suppression function can be approximated using Eq.~\ref{eqn:beta_fit}, for which we fit the  parameters $\alpha$, $\gamma$, and $\Delta$ for fractions in the range of $0.1 \leq f_{\nu_s} \leq 0.6$ \footnote{\CYone{The subhalo suppression functions for the mixed sterile neutrino scenerios considered in this analysis can  be found  \url{https://github.com/chinyitan/mwdm}}.}.


\CYone{We note that there are alternate methods of modeling the phase-space distribution of sterile neutrinos, such as those presented by \citet{Ghiglieri:2015}. While the  phase-space distributions obtained by~\citet{Ghiglieri:2015} are colder than the ones produced by \texttt{sterile-dm} using its default settings, \citet{Boderker:2020} found that both methods are in good agreement with each other when the number of momentum bins used in  \texttt{sterile-dm} is increased to 30,000 to better capture the  resonances in the sterile-neutrino production. For the mixing angles considered in our analysis, we observed that using the transfer function generated from \texttt{sterile-dm} with the increased number of momentum bins does not significantly alter our  mixed sterile neutrino constraints. } 

Fig.~\ref{Figure:sterile_nu_constrained} shows the exclusion region of the fraction of dark matter composed of a 7\,keV sterile neutrino as a function of mixing angle derived from the same galaxy--halo connection model as for the thermal relic WDM scenario. For the mixing angles considered in this work, the constraints on the fraction are roughly constant with $f_{\nu_s} \sim 0.45 (0.55)$ at a posterior ratio of 10:1 (or 20:1). This is consistent with the results for the thermal relic WDM at low mass ($\mwdm\lesssim 2 $~keV). The grey hatched region shows the bound obtained from the maximum allowed lepton asymmetry value from BBN of $L_6\equiv 10^6(n_{\nu_s}-n_{\bar{\nu}_s})/s\le 2500$~\citep{Serpico:2005bc}, which we translate to the $(\fnu,\sin^2(2\theta))$ parameter space using the \texttt{sterile-dm} code. Moreover, sterile neutrinos can decay into an active neutrino and photon, which would be detectable as a monochromatic X-ray line at $E_{\gamma}=m_{\nu_s}/2$. The grey dotted region shows constraints from the non-observation of a line in XMM-Newton data by \citet{Foster:2021ngm}, 
which has been scaled inversely with $f_{\nu_s}$.

\section{Discussion and Conclusion}  \label{sec:summary}

Using the population of MW satellite galaxies detected by  DES and PS1, we put new constraints on MWDM scenarios. We use the CDM subhalo population obtained from  \citet{Mao15} and implemented the MWDM scenario by suppressing the subhalo population using an analytical function, \CYtwo{$\beta(M,\theta_{\rm WDM},\fwdm)$}, derived from the semi-analytical subhalo inference model, \texttt{sashimi} \citep{Hiroshima_2018, Ando_2019, Dekker:2022}. Using the galaxy--halo connection model from \citet{Nadler:MW2, Nadler:MW3}, we estimate the predicted number of MW satellites for different MWDM scenarios and calculate the likelihood of observing the known satellites in DES and PS1 given the MWDM model parameters.

We find that MW satellite galaxy counts constrain the fraction of thermal WDM at $\mwdm \sim 6.6$ keV for $\fwdm = 1.0$ at a posterior ratio threshold of 10:1. As we decrease the WDM mass, our allowed WDM fraction decreases to $\fwdm \sim 0.45$ for $\mwdm \sim 1.5$ keV. These constraints from MW satellite counts extend existing MWDM constraints to much higher masses ($\mwdm > 1$\,keV) compared to previous analysis that used different methods. 
Due to uncertainties in the  galaxy--halo connection model parameters and the limited population of observed MW satellites, \CYtwo{we are not able to constrain MWDM fractions below $\fwdm \lesssim 0.45$ for any $\mwdm$ we consider. However, \citet{Nadler:2024} have demonstrated that with a larger population of fainter satellites, it will be possible to better disentangle the effects of  WDM and the uncertainties in the galaxy--halo connection model parameters. This would therefore allow us to simultaneously achieve better constraints on galaxy--halo connection model parameters and to better distinguish between different dark matter models. }

 In addition to the thermal relic MWDM scenario, we also constrain a mixed dark matter model consisting of CDM and a 7 keV sterile neutrino produced by the Shi-Fuller mechanism. We find that MW satellite galaxy counts constrain a 7 keV sterile neutrino to be a subdominant fraction ($\fnu \lesssim 0.45$) of the dark matter for mixing angles between $11.0  \leq \log_{10}\left ( \sin^2(2\theta) \right) \leq 12.25 $ at a posterior ratio of 10:1. When our constraints are combined with limits from BBN and X-rays, we find that a 7 keV sterile neutrino produced by the Shi-Fuller mechanism is constrained to make up $\lesssim 45\%$ of the dark matter across the range of mixing angles that are commonly considered.

In this work, we relied on only two cosmological zoom-in simulations for our MW subhalo population. This limits our ability to investigate uncertainties in the host halo mass (which changes the normalization of the subhalo mass function and thus the total number of subhalos), host environments and formation histories. 
Recent efforts to simulate a much larger suite of MW-like  hosts in CDM \citep[][]{Nadler:2023,Buch:2024} and alternative dark matter scenarios \citep[][]{Nadler:2024b,An:2024} are expected to better constrain these systematic uncertainties.
In addition, \texttt{sashimi} could be used to generate subhalo populations for a range of host halo masses; however, this would require modifications to incorporate \CYone{ spatial information of the subhalo population  and the effects of the LMC.}

 Our constraints come from the  MW satellite population found in DES and PS1 and the corresponding survey selection functions \citep{Drlica-Wagner:MW1}. However, the known population of MW satellites has grown significantly over the past few years \citep[e.g.,][]{Mau:2020, Cerny:2021, Cerny:2023b, Cerny:2023c, Smith:2023, Homma:2024, Tan:2024}. Furthermore, current and near-future surveys such as the DECam Local Volume Exploration survey \citep[DELVE;][]{Drlica-Wagner:2021, Drlica-Wagner:2022}, the Hyper Suprime-Cam Strategic Survey Program \citep[HSC-SSP;][]{Aihara:2018}, the Ultraviolet Near Infrared Optical Northern Survey \citep[UNIONS;][]{Ibata:2017}, and the Rubin Observatory's Legacy Survey of Space and Time \citep[LSST;][]{Ivezic:2019} are  expected significantly increase the known population of MW satellites in the coming years \citep[e.g.,][]{Hargis:2014, Jethwa:2018, Newton:2018, Nadler:MW2, Manwadkar:2022,Nadler:2024}. This growing population of UFDs will improve our understanding of dark matter and allow for more sensitive searches for potential deviation from CDM.

\begin{acknowledgments}
\CYtwo{The author would like to thank the anonymous referee for the many useful comments that helped us improve this manuscript.} They would also like to thank  Ethan Nadler, Rui An, Vera Gluscevic, and Andrew Benson for the helpful discussions about MWDM scenarios, Jacopo Ghiglieri for his comments about the different sterile neutrino codes,  and Daniel Gilman, Chihway Chang, and Ava Polzin for extended discussion on the statistical formalism used to derive constraints. This work was partially supported by the National Science Foundation (NSF) through AST-2307126. AD is supported by the Kavli Institute for Cosmological physics at the University of Chicago through an endowment from the Kavli Foundation and its founder Fred Kavli. This manuscript has been authored by Fermi Research Alliance, LLC under Contract No.\ DE-AC02-07CH11359 with the U.S. Department of Energy, Office of Science, Office of High Energy Physics.
\end{acknowledgments}



\bibliography{main}

\end{document}